\documentstyle[times,epsf]{mn}
\newcommand{\etal}{{et al}\/.}
\def\figwidth{\linewidth}
\begin{document}
\title[Environments of FRII radio sources]{The environments of FRII
radio sources}
\author[M.J.~Hardcastle \& D.M. Worrall]{M.J.\ Hardcastle and
D.M.\ Worrall \\
Department of Physics, University of Bristol, Tyndall Avenue,
Bristol BS8 1TL}
\maketitle
\begin{abstract}
Using {\it ROSAT} observations, we estimate gas pressures in the
X-ray-emitting medium surrounding 63 FRII radio galaxies and
quasars. We compare these pressures with the internal pressures of the
radio-emitting plasma estimated by assuming minimum energy or
equipartition. In the majority of cases (including 12/13 sources with
modelled, spatially resolved X-ray emission) radio sources appear to
be {\it underpressured} with respect to the external medium,
suggesting that simple minimum-energy arguments underestimate the
sources' internal energy density. We discuss possible departures from
the minimum energy condition and the consequences of our result for
models of the dynamics of radio galaxies, in particular self-similar
models (Kaiser \& Alexander 1997).
\end{abstract}

\section{Introduction}
\label{intro}
Are classical double radio sources (FRIIs; Fanaroff \& Riley 1974)
strongly overpressured with respect to the external medium, or are
they approximately in pressure equilibrium with it? This question is
crucial to our understanding of the dynamics and evolution of radio
sources. It seems clear that FRIIs expand with time along their jet
axis, with the momentum flux supplied by the jet being balanced (on
average) by the ram pressure of the external medium. But, as pointed
out by Scheuer (1974), their lateral expansion, perpendicular to the
jet axis, depends on the difference between the internal (lobe)
pressure and the external (gas) pressure. If the internal pressure is
always much greater than the external pressure (both at all times in
the source's lifetime and at all points along its length), then the
source will expand laterally, at a speed controlled by the pressure
difference; in the limit of strong overpressuring of the lobe the
expansion will be supersonic and an elliptical bow shock will surround
the source [model A of Scheuer (1974)]. On the other hand, if the
internal pressure becomes similar to the pressure in the hot plasma of
the external medium, then the transverse expansion will be subsonic or
will cease entirely, although the supersonic linear expansion will
continue (Scheuer's model C). Scheuer pointed out that the variation
of the external pressure along a source's length could lead to a
situation where the inner parts only are underpressured and contract,
so that eventually buoyancy forces would push the radio lobes away
from the galaxy. (We will discuss this `cocoon crushing' process
further in section \ref{crush}.)

Scheuer (1974) preferred model C, since at that time observations of
radio sources did not show the strong lobe emission that model A
predicts as a result of requiring high internal pressure. Since better
data now show that the lobes are the dominant radio-emitting
components at low frequencies, the objection to model A vanishes, and
many authors take model A to be a good first-order picture of the
evolution of a radio source. But despite the numerous observational
advances of the last 25 years, the question of whether all classical
doubles {\it are} in fact strongly overpressured with respect to their
environments -- that is, whether all classical doubles can be
described by Scheuer's model A -- is still open.

The observation that axial ratios (i.e. the ratio of the length of the
source to its width, by some suitable definition) are similar for
sources of very different lengths (and hence ages) is often taken to
be compelling evidence that the lobes of radio galaxies expand
transversely throughout their lives. If radio sources were confined
transversely throughout their lives, or if they came into pressure
equilibrium with the external medium at a relatively early stage in
their existence, there would be a strong trend for longer sources to
be relatively thinner, which is not observed (Miller \etal\
1985). Recent self-similar models for classical double radio galaxies
(e.g.\ Kaiser \& Alexander 1997) depend on this
transverse expansion, which in turn depends on the internal pressure
of the lobes being much greater than any external pressure from the
radio galaxy's environment.

X-ray observations provide insight into this problem by allowing us to
measure properties of the hot, high-pressure phase of the external
medium. Early investigations based on {\it Einstein} data found
insufficient external pressure to balance the minimum-energy pressure
in a few FRII sources, in support of model A. For example, Arnaud
\etal\ (1984) found that the archetypal classical double 3C\,405
(Cygnus A) had a minimum lobe pressure slightly higher than the
external pressure, and Miller \etal\ (1985) argued that upper limits
on the X-ray emission from a small sample of lower-power FRII sources
implied that they were unconfined by an external atmosphere. However,
lower-power FRI sources inhabit a range of environments, from poor
groups to rich clusters, and it is now well established that the
minimum pressures in {\it their} kpc-scale radio structures are almost
always lower, by an order of magnitude or more, than those in the
X-ray-emitting gas (Morganti \etal\ 1988; Killeen, Bicknell \& Ekers
1988; Feretti \etal\ 1990; Taylor \etal\ 1990; Feretti, Perola \&
Fanti 1992; B\"ohringer \etal\ 1993; Worrall, Birkinshaw \& Cameron
1995; Hardcastle, Birkinshaw \& Worrall 1998b; Worrall \& Birkinshaw
2000a). Subsequent observations with {\it ROSAT} have suggested that
the minimum pressures in some FRIIs are also in fact lower than those
in the X-ray emitting gas (Carilli, Perley \& Harris 1994; Leahy \&
Gizani 1999). This makes it worthwhile to assess the present
observational information on pressures in the external media of a
large sample of FRII sources, and to see what support there is for the
`model A' description of these objects.

We have recently made a study of those sources in the 3CRR sample
(Laing, Riley \& Longair 1983) with pointed {\it ROSAT} observations
(Hardcastle \& Worrall 1999, hereafter paper I).  Slightly over half
of the sample was observed with {\it ROSAT}, and 80 per cent of the
observed sources were detected, giving us a large sample of
observations with good spatial resolution. The FRIIs in the observed
sample span the redshift range from 0.03 to 1.5 and are a good
cross-section of those in 3CRR as a whole. In this paper we discuss
the environments of FRII sources taken from this sample, and the
implications for radio-source models.

Throughout the paper we use a cosmology with $H_0 = 50$ km
s$^{-1}$ Mpc$^{-1}$ and $q_0 = 0$.

\section{Data}

\subsection{Sources with modelled atmospheres}

61 FRII 3CRR sources were observed with {\it ROSAT}, as described in
paper I. Of these, 26 FRII radio galaxies (including 6 broad-line
radio galaxies) and 19 FRII quasars were detected. However, the
majority of the detections were not good enough to allow us to
characterise the spatial distribution of extended emission, either
because of poor statistics or because the X-ray emission was dominated
by a nuclear point source. The FRII objects for which we were able to
fit spatially resolved models to the X-ray data in paper I are
listed in Table \ref{ss}, together with the details of the {\it ROSAT}
observation. In addition to the objects described by Laing \etal\
(1983) as FRIIs, we have included the `jetted double' source 3C\,346
(which though formally an FRI has weak hotspots), and Cygnus A
(3C\,405) which is not a 3CRR object because of its low galactic
latitude.

We have used $\beta$ models to describe the radial distribution
of gas, using the method described by Birkinshaw \& Worrall (1993). In
paper I we fitted only a few, physically reasonable values of $\beta$
to the X-ray data for each source, and quoted the best-fitting
combination of $\beta$ and core radius. Since here we are interested
in determining the full range of uncertainty in the derived pressures,
we adopt the method applied by Worrall \& Birkinshaw (2000b) to
3C\,346 and allow a wide range of values of $\beta$ and core radius in
our fits. For each source we find the best-fit values of the pressure
at the projected inner and outer radii of the radio lobe of
interest. Uncertainties in pressure correspond to $\chi^2 \leq
\chi^2_{\rm min} + 1$ ($\pm 1\sigma$ for one interesting parameter),
where $\beta$, core radius, $\beta$-model normalization and
normalization of a central point source are all free parameters. In
several cases the errors in pressure are highly asymmetrical, allowing
pressures much larger than the best-fit values. These are sources
where the data are centrally peaked, and thought to contain a strong
AGN-related nuclear component, but where the fitting procedure allows,
within the errors, a large fraction of the central peak to be in
high-pressure gas.

An estimate of the hot-gas temperature is necessary to determine the
pressure and its uncertainty. A direct measurement from X-ray data is
used if available. Otherwise, we assume $kT = 1$ keV [based on our
observations of the environments of low-redshift FRI objects; Worrall
\& Birkinshaw (1994, 2000a)] for sources which inhabit low-luminosity
X-ray environments, and a temperature derived from the
temperature-luminosity relation (as discussed in paper I) for the
high-luminosity objects. Where we estimate the temperature, we assume
that the X-ray emission can be described with a Raymond-Smith model
with 0.5 cosmic abundance.

Calculated pressures are tabulated in Table \ref{ss}. The choice of
lobe for pressure estimates is discussed in section \ref{radio-pressures}.

\label{errors}

\subsection{Limits on external pressure}
\label{models}

The results discussed above are for sources in which an extended
environment was detected and separated from the nuclear point source,
where present. There is a risk of bias (in the sense of
selecting the most X-ray luminous environments) if we do not also
consider upper limits from sources without well-characterized
environments, which make up the majority of the {\it ROSAT}-observed
sample. Because we do not have radial-profile information for these
objects, we must adopt a model for the gas distribution and
temperature of the undetected group or cluster. In paper I we assumed
that high-redshift sources without detected extended emission have
environments similar to those of the detected clusters, with $kT = 5$
keV, $\beta = 0.9$ and core radius 150 kpc. This model seems unlikely
to be appropriate for low-redshift FRIIs which, as we know both from
X-ray work (paper I) and optical studies (e.g.\ Prestage \& Peacock
1988), do not typically lie in rich environments. For nearby
FRIIs we have used a model of a typical group-scale atmosphere [based
on our observations of nearby FRIs; Canosa \etal\ (1999), Worrall \&
Birkinshaw (2000a)] which has $kT = 1$ keV, $\beta = 0.35$ and core
radius 40 kpc. We adopt a redshift of 0.3 as the boundary between the
two types of FRII atmosphere. The choice of model parameters does not
affect the derived upper limits on central gas pressure by more than a
factor 3 for typical sources around the boundary redshift.

Of the total of 63 sources, 16 were not detected in the observations
discussed in paper I, and we have determined upper limits on the
central count density by using the upper limits derived for
point-source components in that paper, which were obtained by applying
Poisson statistics to a suitably chosen detection cell. We obtain
limits on the central normalization of a $\beta$-model by considering
how many counts it would contribute to a detection cell, taking the
size of the cell and the PSF of the instrument into account.

For the 14 sources which were detected, but which had too few counts to
allow a convincing radial-profile fit to be carried out, we assume
that the total counts in the source region constitute an upper limit
on the contribution from an extended component in that region, taking
into account the fact that the $\beta$-model will also contribute
some counts to the background region. Choices of source region and
background region are discussed in paper I, but typically for HRI data
an on-source circle of 1 arcmin radius and a background annulus
extending to 2 arcmin were used, while for PSPC data the
corresponding radii were normally 2 and 3 arcmin respectively.

Finally, for the 20 sources which were detected and found to have
radial profiles consistent with the point-spread function (mostly
high-$z$ quasars), we use the technique described in paper I to put an
upper limit on the count rate from an extended component. This
involves simulating observations of a point source and extended
component and finding the count rate at which the extended component
would reliably be detected in the radial profile. For the broad-line
radio galaxy 3C\,390.3 we use the archival PSPC image, rather than the
large number of short HRI observations of Harris, Leighly \& Leahy
(1998) which we discussed in paper I.

These procedures give us limits on central count density, which we
translate into central proton number density and pressure using the
relations given by Birkinshaw \& Worrall (1993). These limits are
tabulated in Table \ref{limits}.

\subsection{Minimum pressures}
\label{radio-pressures}

For each of the observed sources we have used existing radio data to
determine a minimum pressure in one of the radio lobes; we choose the
one which better matches a cylindrical geometry and which is less
affected by compact structure such as jets and hotspots. Minimum
pressures are calculated on the assumption that the radio emission is
synchrotron, and that the only contributions to the internal energy
density come from synchrotron-emitting electrons (and possibly
positrons) and the magnetic field. The minimum energy density which
allows us to obtain the observed synchrotron emissivity can then be
calculated, and the minimum pressure is derived from this.

There are several different possible approaches to calculating this
minimum energy density. Using a number of
simple assumptions, including a power-law distribution of electron
energies with $N(E) {\rm d}E = N_0 E^{-p} {\rm d}E$ between $E = E_{\rm
min}$ and $E_{\rm max}$ and zero elsewhere, the total minimum energy
density in a synchrotron source with cylindrical geometry is
proportional to
\begin{equation}
u_{\rm TOT} \propto \left[(1+\kappa)
{{S\over {\theta_l \theta_r^2 D_{\rm L}}}
}I\right]^{4\over p+5}
\label{mine}
\end{equation}
where $S$ is the observed radio flux, $\theta_l$ and
$\theta_r$ are the observed angular length and radius respectively,
$D_{\rm L}$ is the luminosity distance to the source,
and $\kappa$ is the
ratio between the energy densities in non-radiating and radiating
particles; $I$ is a function of the energy range of the electron power
law, defined as
\[
I = \left\{ \begin{array}{ll}\ln(E_{\rm max}/E_{\rm min})&p=2\\
{1\over{2-p}} \left[E^{(2-p)}_{\rm max}-E^{(2-p)}_{\rm min}\right]&p\neq
2\\
\end{array}\right .
\]
There is a relatively weak dependence of the minimum
energy on parameters such the source dimensions and $\kappa$, and an
extremely weak dependence on the energy range used.

To perform minimum-energy calculations for our sources we use computer
code which performs the synchrotron emissivity and electron energy
integrals numerically, thus allowing us to use more complex electron
energy spectra. Normally we assume that the electron energy spectrum
is a power law with energy index $p$ of 2 between $E_{\rm min} = 5
\times 10^6$ eV ($\gamma_{\rm min} = 10$) and $E_{\rm max} = 5 \times
10^{10}$ eV ($\gamma_{\rm max} = 10^5$); this is the standard energy
index derived for first-order Fermi acceleration at strong shocks. For
our flux measurements we have used the lowest-frequency available
radio data of good quality, so as to minimise the effect of any
age-related steepening in the electron spectrum (which is most marked
at high electron energies and so high radio frequencies) and to reduce
the contribution of hotspots, which have flat spectra. Where two or
more radio frequencies are available, we have allowed the electron
energy index to steepen to 3 at a best-fit energy, to account roughly
for the effects of synchrotron ageing. We set $\kappa$ to zero and
approximate the lobes as uniform cylinders to derive an average `lobe
pressure'. [We expect that the pressure is reasonably constant
throughout the lobes of an FRII radio source, except at or close to
the hotspot, because of the high expected sound speed in the
radio-emitting plasma, as argued by Kaiser \& Alexander (1997).] We
assume that the sources are in the plane of the sky; this means
(equation \ref{mine}) that we overestimate the minimum pressure by a
factor of $\sim (\sin \theta)^{-4/7}$, where $\theta$ is the angle to
the line of sight, but this factor is small compared to the other
uncertainties in the calculation unless $\theta$ is very small (see
section \ref{projection}).

For some of the sources without well-characterised X-ray environments,
radio maps were not available to us in digital form, and we used total
178-MHz flux densities from Laing \etal\ (1983),
corrected to the Baars \etal\ (1977) flux scale, and total source size
from published maps to estimate an average minimum pressure. Where the
published maps were not good enough to give us an estimate of the
width of the source we assumed an axial ratio (ratio of total length to total
width) of 4.5.

Minimum pressures are tabulated in Tables \ref{limits} and \ref{pressure}.

\begin{table*}
\caption{X-ray pressure measurements for sources with modelled X-ray environments}
\label{ss}
\begin{tabular}{lrrrrrrr}
\hline
Source&$z$&Livetime&$kT$ used&$r_{\rm
min}$&$r_{\rm max}$&$p(r_{\rm min})$&$p(r_{\rm max})$\\
&&(s)&(keV)&(arcsec)&(arcsec)&(Pa)&(Pa)\\

\hline
3C\,98&0.0306 &41047 (H)&1.0&14&160&$3.3^{+4.4}_{-1.5} \times 10^{-13}$&$4.2^{+7.3}_{-2.9} \times 10^{-14}$\\
3C\,123&0.2177 &28801 (H)&3.6&7&18&$1.2^{+0.3}_{-0.3} \times 10^{-11}$&$6.0^{+0.6}_{-0.7} \times 10^{-12}$\\
3C\,215&0.411 &86442 (H)&4.0&0&25&$8.8^{+18.2}_{-3.9} \times 10^{-12}$&$2.5^{+0.3}_{-0.4} \times 10^{-12}$\\
3C\,219&0.1744 &4206 (P)&1.0&0&92&$9.8^{+10300}_{-3.4} \times 10^{-13}$&$5.4^{+5.9}_{-3.9} \times 10^{-14}$\\
3C\,220.1&0.61 &36226 (H)&5.6&6&18&$2.0^{+1.1}_{-0.6} \times 10^{-11}$&$8.4^{+0.4}_{-1.5} \times 10^{-12}$\\
3C\,254&0.734 &15570 (P)&7.7&0&10&$7.7^{+2.2}_{-1.2} \times 10^{-12}$&$7.4^{+2.1}_{-1.2} \times 10^{-12}$\\
3C\,275.1&0.557 &25158 (H)&4.8&0&12&$7.2^{+4.8}_{-2.4} \times 10^{-12}$&$6.2^{+2.0}_{-1.8} \times 10^{-12}$\\
3C\,280&0.996 &46619 (P)&5.0&0&12&$8.3^{+180.3}_{-4.0} \times 10^{-13}$&$8.2^{+35.0}_{-3.9} \times 10^{-13}$\\
3C\,295&0.4614 &29292 (H)&4.4&0&2&$6.0^{+1.8}_{-1.2} \times 10^{-11}$&$5.7^{+1.4}_{-1.1} \times 10^{-11}$\\
3C\,334&0.555 &27909 (H)&5.4&13&34&$6.7^{+2.6}_{-2.9} \times 10^{-12}$&$1.7^{+0.5}_{-1.3} \times 10^{-12}$\\
3C\,346&0.162 &16981 (P)&1.9&0&10&$1.1^{+1.0}_{-0.5} \times 10^{-12}$&$1.1^{+0.9}_{-0.4} \times 10^{-12}$\\
3C\,388&0.0908 &52674 (H)&3.1&0&28&$1.9^{+0.5}_{-0.4} \times 10^{-11}$&$4.9^{+0.2}_{-0.2} \times 10^{-12}$\\
3C\,405&0.0565&9127 (P)&7.3&0&70&$2.0^{+0.1}_{-0.1} \times 10^{-10}$&$1.8^{+0.1}_{-0.1} \times 10^{-11}$\\
\hline
\end{tabular}
\vskip 10pt
\begin{minipage}{\linewidth}
Redshifts are taken from Laing \etal\ (1983). An H in column 3
indicates that the data come from the {\it ROSAT} HRI; a P indicates
the PSPC. Pressures are calculated at the radii $r_{\rm min}$
and $r_{\max}$, which correspond to the (projected) minimum and
maximum radii sampled by the radio lobes. Errors on the pressures are
computed as described in the text. Temperatures are estimated from the
temperature-luminosity relation or set to 1 keV (see the text) except
for those sources where temperature measurements exist: these are
3C\,220.1 (Ota \etal\ 2000), 3C\,295 (Harris \etal\ 2000), 3C\,346
(Worrall \& Birkinshaw, 2000b) and 3C\,405 (Ueno
\etal\ 1994).
\end{minipage}
\end{table*}

\begin{table*}
\caption{Radio measurements and upper limits on central thermal
pressure for sources without modelled X-ray environments}
\label{limits}
\begin{tabular}{lrrrrrrrrl}
\hline
Source&Lobe&Flux&Freq.&Length&Width&$p_{\rm min}$&$z$&$p_0$&Ref.\\
&&(Jy)&(GHz)&(arcsec)&(arcsec)&(Pa)&&(Pa)\\
\hline
3C\,13&Whole &13.08 &0.178 &25 &-- &$3.4 \times 10^{-12}$ &  1.351 &$< 1.9 \times 10^{-11}$&1\\
3C\,20*& E&5.2 &1.41 &29 &26 &$3.4 \times 10^{-13}$ &  0.174 &$< 1.0 \times 10^{-11}$&2,3\\
3C\,33& N (part) & 0.72 & 1.4 &68 &101 &$2 \times 10^{-14}$ &  0.0595&$< 8 \times 10^{-13}$&4\\
3C\,33.1 & E &1.69& 1.53 &88 &60 &$4 \times 10^{-14}$ &  0.181 &$<3.5 \times 10^{-12}$&5\\
3C\,47*&N &1.26 &1.65 &27 &20 &$3.0 \times 10^{-13}$ &  0.425 &$< 2.0 \times 10^{-11}$&6,7\\
3C\,61.1&S (part) &1.24 &1.48 &35 &50 &$5.9 \times 10^{-14}$ &  0.186 &$< 6 \times 10^{-12}$&4\\
3C\,67&N&0.79 &1.67 &1.7 &1.0 &$2.3 \times 10^{-11}$ &  0.3102 &$< 9.1 \times 10^{-12}$&8\\
3C\,79*&E (part) &1.20 &1.45 &36 &14 &$3.7 \times 10^{-13}$ &  0.2559 &$< 2.0 \times 10^{-11}$&9,3\\
3C\,171*&E &1.75 &1.44 &20 &7.4 &$1.4 \times 10^{-12}$ &  0.2384 &$< 6 \times 10^{-12}$&2,3\\
3C\,181&Whole &15.81 &0.178 &6.7 &1.3 &$5 \times 10^{-11}$ &  1.382 &$< 7.0 \times 10^{-11}$&10\\
3C\,192*&E &2.7 &1.41 &89 &55 &$1.1 \times 10^{-13}$ &  0.0598 &$< 4 \times 10^{-12}$&11,12\\
3C\,196&N &0.52 &14.96 &4.9 &2.3 &$1.1 \times 10^{-11}$ &  0.871 &$< 2.5 \times 10^{-11}$&2,13\\
3C\,204&E &0.122 &4.9 &7.3 &2.7 &$2.7 \times 10^{-12}$ &  1.112 &$< 7.2 \times 10^{-11}$&7\\
3C\,207&Whole &14.82 &0.178 &18 &12 &$1.2 \times 10^{-12}$ &  0.684 &$< 3.3 \times 10^{-11}$&14\\
3C\,208&E &0.400 &4.9 &5.5 &2 &$8.8 \times 10^{-12}$ &  1.109 &$< 4.4 \times 10^{-11}$&7\\
3C\,212&Whole &16.46 &0.178 &11 &-- &$1.2 \times 10^{-11}$ &  1.049 &$< 3.8 \times 10^{-11}$&15\\
3C\,220.3&Whole &17.11 &0.178 &11 &4 &$5.3 \times 10^{-12}$ &  0.685 &$< 1.7 \times 10^{-11}$&16\\
3C\,223& S & 1.53 &1.50 &145 &55 &$1 \times 10^{-14}$ &  0.1368 &$< 1 \times 10^{-12}$&4\\
4C\,73.08&E&4.56 & 0.61 &410 &298 &$4 \times 10^{-15}$& 0.0581 &$< 6.6 \times 10^{-13}$&5\\
3C\,236& N&1.95 &0.61 &781 &229 &$2 \times 10^{-15}$ &  0.0989 &$< 5.5 \times 10^{-13}$&17\\
3C\,241&Whole &12.64&0.178 &0.9 &-- & $1.0 \times 10^{-9}$  &  1.617 &$< 4.9 \times 10^{-11}$&18\\
3C\,245&E &0.134 &4.89 &2.9 &1.9 &$1.0 \times 10^{-11}$ &  1.029 &$< 5.1 \times 10^{-11}$&19\\
3C\,247*&S &1.76 &1.46 &8.6 &3.5 &$6.3 \times 10^{-12}$ &  0.7489 &$< 3.5 \times 10^{-11}$&19\\
3C\,249.1&W &1.45 &1.42 &15 &11 &$5.3 \times 10^{-13}$ &  0.311 &$< 2.5 \times 10^{-11}$&5\\
3C\,263&E &0.739 &4.9 &12 &8 &$1.6 \times 10^{-12}$  &  0.6563 &$< 2.0 \times 10^{-11}$&7\\
3C\,263.1&N &0.326 &4.89 &3.1 &1.6 &$1.4 \times 10^{-11}$ &  0.824 &$< 2.4 \times 10^{-11}$&19\\
3C\,266&Whole &12.19 &0.178 &5.4 &0.8 &$8.1 \times 10^{-11}$ &  1.2750 &$< 1.9 \times 10^{-11}$&1\\
3C\,268.3&S &0.381 &4.99 &0.7 &0.4 &$9.6 \times 10^{-11}$ &  0.371 &$< 4.0 \times 10^{-11}$&20\\
3C\,268.4*&N &0.386 &1.47 &3.0 &1.6 &$2.1 \times 10^{-11}$ &  1.400 &$< 9.5 \times 10^{-11}$&19\\
3C\,270.1*&S &1.96 &1.46 &4 &2 &$4.1 \times 10^{-11}$ &  1.519 &$< 6.4 \times 10^{-11}$&15,19\\
3C\,277.2&W&1.54 &1.41 &16 &6.1 &$1.6 \times 10^{-12}$ &  0.766 &$<
6.9 \times 10^{-12}$&21,22\\
3C\,284*&W &0.803 &1.53 &105 &23 &$9.5 \times 10^{-14}$ &  0.2394 &$< 1.2 \times 10^{-12}$&23,3\\
3C\,289&E &1.43 &1.46 &5.5 &2.7 &$8.1 \times 10^{-12}$ &  0.9674 &$< 1.2 \times 10^{-11}$&19\\
3C\,294*&S &0.478&1.46&9.5 &4.2 & $6.5 \times 10^{-12}$&  1.78 &$< 2.6 \times 10^{-11}$&19\\
3C\,299&E &2.69 &1.53 &2.6 &0.6 &$2.6 \times 10^{-11}$ &  0.367 &$< 2.0 \times 10^{-11}$&5\\
3C\,303&E &0.39 &1.45 &20 &18 &$1.4 \times 10^{-13}$ &  0.141 &$< 1.5 \times 10^{-12}$&4\\
3C\,318&Whole &13.41 &0.178 &0.93 &0.13 &$2.1 \times 10^{-9}$ &  1.574 &$< 6.6 \times 10^{-11}$&24\\
3C\,324&Whole &17.22 &0.178 &9.4 &2 &$2.2 \times 10^{-11}$ &  1.2063 &$< 2.2 \times 10^{-11}$&25\\
3C\,326&N &1.85 &1.40 &374 &289 &$3 \times 10^{-15}$ &  0.0895 &$< 1.3 \times 10^{-12}$&5\\
3C\,325&Whole &17.00 &0.178 &16.8 &2.2 &$1.0 \times 10^{-11}$ &  0.86 &$< 2.6 \times 10^{-11}$&26\\
3C\,330&Whole &30.30 &0.178 &63 &-- &$6.7 \times 10^{-13}$ &  0.5490 &$< 1.2 \times 10^{-11}$&26\\
3C\,343.1&Whole &12.54 &0.178 &0.4 &0.12 &$2.0 \times 10^{-9}$ &  0.750 &$< 1.6 \times 10^{-11}$&18\\
3C\,351&S &0.33 &1.42 &27 &7 &$5.6 \times 10^{-14}$ &  0.371 &$< 1.2 \times 10^{-11}$&4\\
3C\,356&Whole &12.32 &0.178 &76 &12 &$6.7 \times 10^{-13}$ &  1.079 &$< 1.3 \times 10^{-11}$&27\\
4C\,16.49&Whole &11.45 &0.178 &18 &2 &$1.3 \times 10^{-11}$ &  1.296 &$< 6.3 \times 10^{-11}$&28\\
3C\,368&Whole &15.04 &0.178 &8.8 &1.6 &$2.6 \times 10^{-11}$ &  1.132 &$< 1.5 \times 10^{-11}$&25\\
3C\,382&S &2.22 &1.45 &143 &67 &$1.7 \times 10^{-14}$ &  0.0578 &$< 2 \times 10^{-12}$&4\\
3C\,390.3&N (part)&1.62 &1.45 &88 &71 &$1.4 \times 10^{-13}$ &0.0569&$< 5.6 \times 10^{-12}$&5\\
3C\,433&S &1.83 &8.47 &32 &14 &$6.3 \times 10^{-13}$ &  0.1016 &$< 1.6 \times 10^{-12}$&29\\
3C\,455&Whole &13.95 &0.178 &4.3 &1.2 &$3.3 \times 10^{-11}$ &  0.5427&$< 8.0 \times 10^{-11}$&14\\
\hline
\end{tabular}
\vskip 10pt
\begin{minipage}{\linewidth}
The radial profile and temperature models used are determined by the
redshift (see the text). For sources marked with an asterisk we have
multi-frequency radio data and have fit a model spectrum to the
source, as described in section \ref{radio-pressures}. References for radio
maps and size measurements are: (1) Best, Longair \& R\"ottgering
(1997); (2) Laing (unpublished) (3) Hardcastle \etal\ (1997); (4)
Leahy \& Perley (1991); (5) Leahy, Bridle \& Strom (1998); (6) Leahy
(1996); (7) Bridle \etal\ (1994); (8) Katz-Stone \& Rudnick (1997);
(9) Spangler, Myers \& Pogge (1984); (10) Mantovani \etal\ (1994);
(11) Laing, published in Baum \etal\ (1988); (12) Leahy \etal\ (1997);
(13) Brown (1990); (14) Bogers \etal\ (1994); (15) Akujor \etal\
(1991); (16) Jenkins, Pooley \& Riley (1977); (17) Mack \etal\ (1997);
(18) Fanti \etal\ (1985); (19) Liu, Pooley \& Riley (1992); (20)
L\"udke \etal\ (1998); (21) Alexander \& Leahy (1987) (22) Pedelty
\etal\ (1989); (23) Leahy, Pooley \& Riley (1986); (24) Spencer \etal\
(1991); (25) Best \etal\ (1998); (26) Fernini, Burns \& Perley (1997);
(27) Fernini \etal\ (1993); (28) Lonsdale, Barthel \& Miley (1993);
(29) Black \etal\ (1992). Maps of $z<0.5$ objects were mostly obtained
from Leahy \etal\ (1998). Redshifts are taken from Laing \etal\ (1983)
except for that of 3C\,318, which is taken from Willott, Rawlings \&
Jarvis (1999).
\end{minipage}
\end{table*}

\begin{table*}
\caption{Radio measurements for the sources with modelled X-ray environments}
\label{pressure}
\begin{tabular}{lrrrrrrrrrrr}
\hline
Source&Lobe&Flux&Freq.&Length&Width&$p_{\rm min}$\\
&&(Jy)&(GHz)&(arcsec)&(arcsec)&(Pa)\\
\hline
3C\,98&S&1.42&8.35&145&60&$7 \times 10^{-14}$\\
3C\,123*&N&11.3&1.43&14&5&$5.8 \times 10^{-12}$\\
3C\,215*&N&0.203&4.84&25&18&$3.4 \times 10^{-13}$\\
3C\,219*&S&3.70&1.52&92&46&$1.5 \times 10^{-13}$\\
3C\,220.1&W&1.20&1.40&12&7&$1.2 \times 10^{-12}$\\
3C\,254&W&1.08&1.40&10&4&$2.5 \times 10^{-12}$\\
3C\,275.1&N&0.205&8.46&12&6&$8.2 \times 10^{-13}$\\
3C\,280&W&0.448&14.96&11&6&$2.3 \times 10^{-12}$&\\
3C\,295*&N&1.48&8.56&2.3&1.2&$4.2 \times 10^{-11}$\\
3C\,334*&N&0.276&4.84&26&13&$6.0 \times 10^{-13}$\\
3C\,346&S&1.19&1.53&7&9&$1.0 \times 10^{-12}$\\
3C\,388&N&2.93&1.39&28&18&$4.6 \times 10^{-13}$\\
3C\,405*&S&205.6&4.53&48&25&$7 \times 10^{-12}$\\
\hline
\end{tabular}
\vskip 10pt
\begin{minipage}{\linewidth}
Sources of radio maps and size/flux measurements are as follows:
3C\,98, Leahy \etal\ (1997); 3C\,123, Laing (unpublished) and
Hardcastle \etal\ (1997); 3C\,215, Bridle \etal\ (1994); 3C\,219,
Clarke \etal\ (1992); 3C220.1, Harvanek \& Hardcastle (1998) and Burns
\etal\ (1984); 3C\,254, Liu \etal\ (1992) and FIRST survey data;
3C\,275.1, Gilbert \etal\ (in prep.); 3C\, 280, Laing (unpublished)
and Liu \etal\ (1992); 3C\,295, Perley \& Taylor (1991) and Cotton
(unpublished); 3C\,334, Bridle \etal\ (1994); 3C\,346, Leahy, Bridle
\& Strom (1998); 3C\,388, Roettiger \etal\ (1994); 3C\,405, Carilli
\etal\ (1991). Electronic images for $z<0.5$ objects were mostly
obtained from Leahy \etal\ (1998). For sources marked with an asterisk
we have multi-frequency radio data and have fit a model spectrum to
the source, as described in section \ref{radio-pressures}. The
minimum-energy fit to 3C\,123's N lobe is taken from Looney \&
Hardcastle (2000).
\end{minipage}
\end{table*}

\section{Why do radio sources appear underpressured?}
\label{small}

\begin{figure}
\caption{Plot of central thermal pressure against minimum
pressure for the X-ray observed sources. The solid line shows equality
of internal and external pressure. Most sources lie above it. Crosses
denote sources with $z<0.3$, and stars sources with $z>0.3$. Arrows
denote upper limits on thermal pressure; upper limits are plotted in
light grey.}
\label{prs}
\epsfxsize
\figwidth\epsfbox{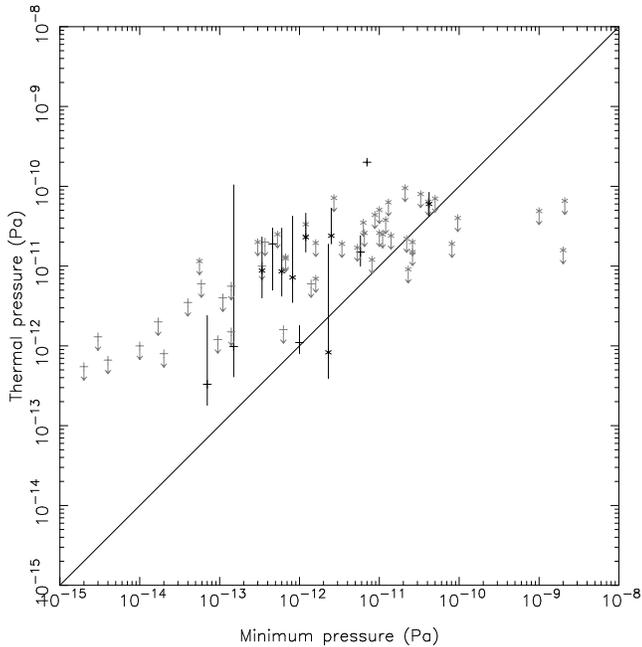}
\end{figure}

Tables \ref{ss}, \ref{limits} and \ref{pressure} and Fig.\ \ref{prs}
show that, with very few exceptions, the minimum pressures of the
radio lobes are well below the measured pressures or upper limits on
pressure in the central part of the hot-gas component of the
environment. In the majority of the detected sources in Tables \ref{ss} and
\ref{pressure} the cluster pressure is higher than the lobe pressure
even at the far ends of the sources, where the cluster pressure is
lowest.

\begin{figure}
\caption{The ratio between central thermal pressure and minimum
pressure as a function of lobe length. (Lengths plotted for the
sources for which minimum pressure is calculated for the whole source
are half the source length.) The solid line shows equality of internal
and external pressure. Crosses denote sources with $z<0.3$, and stars
sources with $z>0.3$. Arrows denote upper limits on thermal pressure;
upper limits are plotted in light grey. The correlation seen in this
figure arises mainly because of a strong anticorrelation between
minimum pressure and source size.}
\label{prl}
\epsfxsize
\figwidth\epsfbox{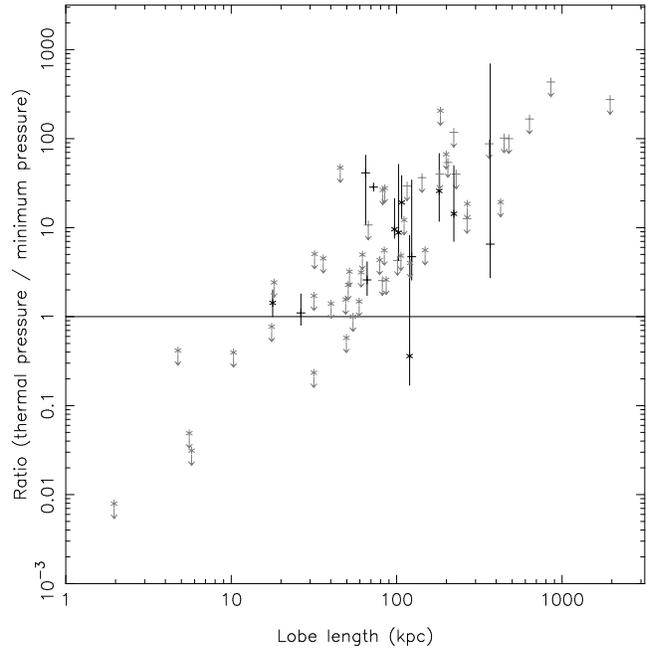}
\end{figure}

As shown in Fig.\ \ref{prl}, the few sources with minimum
pressures higher than the limits on their central thermal pressures are all
small objects, with lobe lengths less than the (assumed) core
radii. This seems to be mainly because of a strong anticorrelation
between minimum pressure and source size; sources which are small
($\la 10$ kpc) and luminous enough to be in the 3CRR catalogue
naturally have very high minimum pressures. This overpressuring with
respect to the external medium is what we would
expect to find if the small sources are young, a point we
return to briefly in section \ref{conseq}. In this section we shall
concentrate on the more typical sources with linear sizes $\ga 10$ kpc.

There are several reasons why the true radio-lobe pressure may be
closer to the thermal pressure than implied by Fig.\ \ref{prs}. In
this section of the paper we consider them in turn. Our choice of $p=2$
determines the index $4/7$ that appears in many of the approximate
relations we quote (see equation \ref{mine}).

\subsection{X-ray uncertainties}

The thermal pressures depend directly on the X-ray emission. The
accuracy of our pressure estimates thus depends on the quality of
our data, and in particular on whether the source has spatially
resolved, modelled structure, whether it is detected, and whether it
is dominated by a non-thermal point source, as follows.

\begin{itemize}
\item {\it Sources with modelled emission.} Figures \ref{prs} and
\ref{prl} show that the errors on the thermal pressures in these
sources do not in general allow thermal pressures to lie below the
minimum pressures.
\item {\it Detected sources with too few counts to model.}
14 of the upper limits on pressure are based on X-ray detections which had
too few counts to allow modelling of the sources' spatial structure. 
These are upper limits only in the sense that there is
an unknown contribution from non-thermal emission. All of these
sources (except 3C\,236) have weak radio cores, so that
from the observed correlation between radio and X-ray nuclear
emission (paper I) we would not expect a strong
contribution to the X-ray emission from the active nucleus; the
detected X-ray emission will come almost
entirely from hot gas. The estimated limits on pressure should then
be very close to the true values for these sources.
\item{\it Non-detections.} The majority of the 16 non-detected sources
have short or off-axis observations. Since we are aware of no bias in
the {\it ROSAT} observations in the sense that more sensitive
observations were made for sources with suspected richer cluster
environments, the non-detections are consistent with having been drawn
from the same population as the detections. It is nevertheless
possible that the true values for the thermal pressures in these
sources lie a long way below our limits.
\item{\it Detections with dominant point source.} Our ability to
detect atmospheres in these 20 sources is limited by the strength of
the central component (and by uncertainties in the HRI PSF) rather
than by sensitivity. Obviously here, as for the non-detections, we
cannot rule out the possibility that in some or all cases the true
pressures lie a long way below the upper limits.
\end{itemize}

Some of the FRIIs may have much lower thermal pressures than the upper
limits that appear in Table \ref{limits} and Figures \ref{prs} and
\ref{prl}. But without any evidence to the contrary, the simplest
picture is that most FRIIs with linear sizes $\ga 10$ kpc are similar
to our detected sources and so are underpressured.
\subsection{Cosmology}

Our choice of cosmology does not significantly affect the relationship
between radio and thermal pressures. If the Hubble parameter is $50h$
km s$^{-1}$ Mpc$^{-1}$, then the minimum pressure in the radio lobes
goes approximately as $h^{4\over 7}$ (equation \ref{mine}), while the
thermal pressure goes as $h^{1\over 2}$ for a detected X-ray
environment whose angular scale we know; no sensible change in $h$ can
eliminate the pressure differences. For upper limits, we have chosen a
fixed linear size, and so it might seem that the estimated upper limit
on thermal pressure is independent of $h$. However, since the linear
size we choose is taken from observations on the assumption of $h=1$,
the upper limits on thermal pressure go as $h^{1/2}$ in this case too.

\subsection{Effects of projection}
\label{projection}

\begin{figure}
\caption{Contours of $\cal K$, the ratio between projection-corrected
and inferred ratios of thermal to radio pressure, as a function of $R$
and $\theta$. The figure is calculated for $\beta = 0.67$; similar results are
obtained with $\beta = 0.5$ and $\beta = 0.9$.}
\label{ratiocont}
\epsfxsize \figwidth
\epsfbox{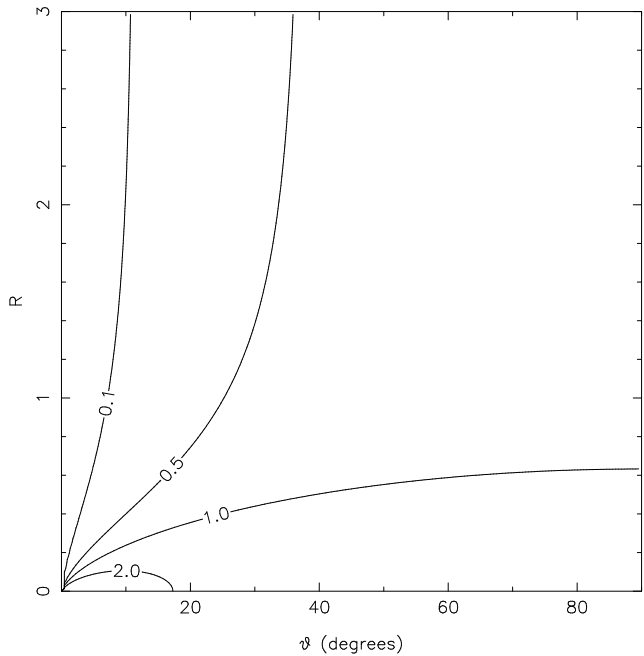}
\end{figure}

Projection affects the ratio of radio to thermal pressure according to
the approximate relation
\[
{p_{\rm th}\over{p_{\rm r}}} = {\cal K}\left({p'_{\rm th}\over{p'_{\rm
r}}}\right) = (\sin \theta)^{-4/7} \left({1 +
R^2}\over{1 + R^2/\sin^2 \theta}\right )^{3\beta/2} \left({p'_{\rm th}\over{p'_{\rm
r}}}\right)
\]
where $p'_{\rm th}$ and $p'_{\rm r}$ are the originally inferred
thermal and radio pressures, $\theta$ is the angle of the radio
structure to the line of sight and $R$ is the ratio between the
apparent (projected) distance along the radio galaxy and the core
radius of the cluster (Birkinshaw \& Worrall 1993). Depending on the
values of $R$ and $\theta$, projection can either decrease or increase
the pressure ratio. Fig.\ \ref{ratiocont} shows contours of values of
this relation as a function of $R$ and $\theta$. It will be seen that
for small $R$ the assumption of no projection causes us to {\it
underestimate} the ratio of thermal to minimum pressure (${\cal
K} > 1$). This may be important, for example, in the case of 3C\,346,
which appears to have similar X-ray and radio pressures, but whose
radio structure (prominent core, bright one-sided jet, small linear
size) is suggestive of projection effects, and which lies in a cluster with a
large core radius, so that $R$ is small even at the end of the source
(see also Worrall \& Birkinshaw 2000b). It may also be important
for some of the small sources in Table \ref{limits} whose
inferred minimum lobe pressures are close to or exceed the upper
limits on central pressure. For most sources in our sample $R$ is zero
for the inner pressure measurement, and therefore projection can only
cause the corrected minimum pressure to be smaller than the estimated
minimum pressure. For larger $R$ and small $\theta$ Fig.\
\ref{ratiocont} shows that we overestimate the ratio of thermal
pressure to minimum pressure, but this effect only becomes large
(${\cal K} < 0.1$) for very small angles to the line of sight ($\theta
\la 10^\circ$); few, if any, of the observed sources are likely to be
this strongly projected.  Therefore it is unlikely that projection
alone can be responsible for the discrepancy between radio and thermal
pressures.

\subsection{Radio-related contributions to the X-ray emission}

We have assumed that all the extended X-ray emission in the detected
sources can be attributed to the hot intra-cluster medium. Brunetti,
Setti \& Comastri (1997) argue that there may be a significant
contribution to the extended X-ray emission of radio sources from
inverse-Compton scattering of the IR--optical photons from a central
quasar by the low-energy electron population in the radio lobes. In
addition, there is necessarily inverse-Compton emission from
scattering of cosmic microwave-background (CMB) photons. If these processes
are important in our objects, then they will cause us to overestimate
the thermal pressures in the cluster environment. For some
well-studied, low-redshift objects in our sample, it is clear from the
observations that the X-ray emission is dominated by cluster emission,
since it is approximately radially symmetrical and extends further
than the radio lobes. This is true, for example, of 3C\,123, 3C\,295,
3C\,346, 3C\,388 and 3C\,405 in Table \ref{ss}. For the distant object
3C\,220.1, the ASCA observations of Ota \etal\ (2000), in which an
iron line is detected, seem to confirm the conclusion of Hardcastle,
Lawrence \& Worrall (1998) that the extended emission is
cluster-related.  On the other hand, Brunetti \etal\ (1999) suggest in
the case of 3C\,219 that a large fraction of the extended emission in
an HRI image is inverse-Compton in origin, which, if true, would mean
that the pressures we estimate for the thermal emission in this source
are too high by up to an order of magnitude. For the other sources,
most of which are at high redshift, the spatial resolution and
sensitivity of existing X-ray data is inadequate to distinguish
between the two models, and we must await planned {\it Chandra}
observations. But on the balance of the evidence so far, and given the
optical evidence pointing to the existence of clusters around
high-redshift objects, we feel justified in our assumption that the
extended emission is dominated by a thermal intra-cluster component.

\subsection{Temperature assumptions}

\begin{figure}
\caption{Central thermal cluster pressure as a function of assumed
temperature for a source of fixed {\it ROSAT} HRI count rate (the
calculation is done for the $z=0.61$ source 3C\,220.1). The broad {\it
ROSAT} passband means that pressure is approximately linearly
dependent on temperature in the range 0.5--10 keV, but then starts to
rise again as the source moves below the {\it ROSAT}
passband. Temperatures between $\sim 0.05$ and $\sim 1$ keV would
bring the central pressure in this source below the minimum pressure
in the lobes (shown by the dashed line). The cross marks our adopted
temperature of 5.6 keV for 3C\,220.1.}
\label{temp}
\epsfxsize \figwidth
\epsfbox{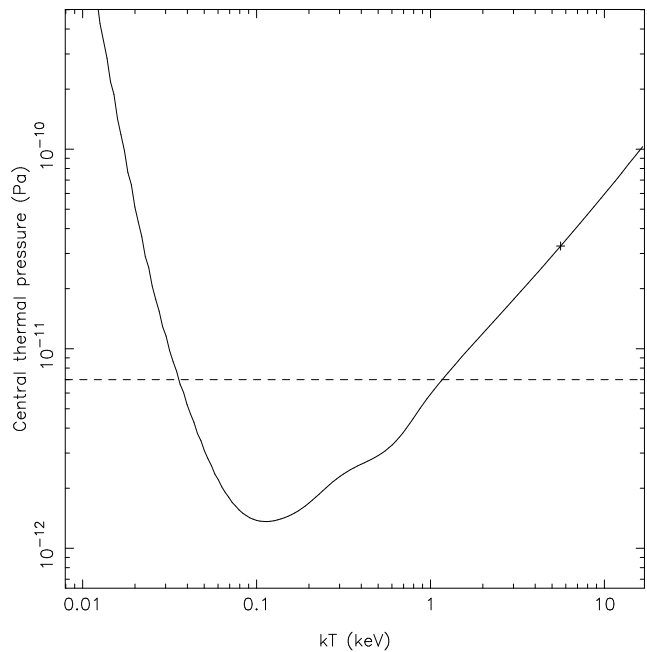}
\end{figure}

In many cases, including all the upper limits, we have no adequate
measurement of the temperature of the hot X-ray emitting gas. The
estimates for about half the detected sources are based on the
temperature-luminosity relationship of David \etal\ (1993), which
appears to hold out to redshifts $\sim 0.5$ or even greater
(Mushotzky \& Scharf 1997, Donahue \etal\ 1998), comparable with the
highest redshifts where we have been able to separate nuclear and
extended emission. There is a good deal of scatter in the
temperature-luminosity relation and so these results are uncertain by
perhaps a factor of 2. As shown in Figure \ref{temp}, inferred pressure
is approximately linearly dependent on assumed temperature for $kT >
0.5$ keV, and so to account for the discrepancy between thermal and
minimum radio pressures the estimated temperatures would need to be
systematically high by a factor 5--10, which seems unlikely.

Some of the sources we have considered may well contain cooling flows,
so that the whole idea of a uniform `cluster temperature' may be
misleading. In general, cooling flows will only affect the innermost
regions of the cluster and are not relevant on scales comparable to
the linear size of the radio source, but they may cause us to
overestimate central cluster pressures; for example, the pressure in
the central bin of the deprojection analysis by Reynolds \& Fabian
(1996) of the cooling flow in the Cygnus A cluster is a factor 3 lower
than the central pressure quoted in Table \ref{pressure} (though still
much higher than the minimum pressure in the lobes) while the pressure
at 70 arcsec from the core is similar to our value. More
observations are required to measure the influence of possible cooling
flows in the cluster environments of these sources.

\subsection{Electron spectrum assumptions}

In calculating minimum energies we have assumed a fixed value ($p=2$)
for the power-law index of the electron energy distribution at low
energies, corresponding to a low-frequency spectral index $\alpha$
(the `injection index') of 0.5 ($\alpha = (p-1)/2$). This should be
realistic if the electrons in the lobes were accelerated at
non-relativistic strong shocks. The energy density in electrons, and
therefore the radio-related pressure, depends strongly on our choice
of $p$ (equation \ref{mine}). For sources which are described by a
simple power law (i.e. with no spectral break) we obtain minimum
pressures which are a factor $\sim 2$--$4$ higher for $p=2.5$ and
$\sim 4$--$16$ higher for $p=3$ (the exact value depends on the
frequency of flux measurement). If we were to adopt $p=3$ for all
sources we would obtain minimum pressures close to or exceeding the
thermal values in many cases. However, such a high value of $p$ is
inconsistent with particle acceleration models and with observations of
the low-energy spectral indices of radio sources, where $\alpha$ is
generally significantly less than 1.

At frequencies $>1$ GHz, corresponding to the majority of the radio
observations we have used, $\alpha$ is typically greater than 0.5 in
the lobes. The steeper spectral index is conventionally attributed to
spectral ageing effects. Where possible we have made a rough
correction for these steeper values of $\alpha$ by using more than one
frequency and fitting a high-energy break in the electron
spectrum. (Sources where this has been done are marked with an
asterisk in Tables \ref{limits} and \ref{pressure}.) For sources with
insufficient spectral information we have neglected spectral ageing,
and so our synchrotron flux may underestimate the normalization of the
low-energy electron spectrum, causing us to underestimate the minimum
pressure. However, the underestimation is at most a factor 2 in a
typical source.

We discuss the effects of varying the low-energy cutoff of the
electron spectrum in section \ref{nonr}.

\subsection{The minimum energy/equipartition assumption}
\label{icobs}

There is little strong justification for the assumption that the radio
lobes are near to their minimum pressures, or, roughly equivalently,
that there is equal energy density in radiating electrons and magnetic
fields. To provide an order of magnitude increase in radio pressure
the magnetic field strength in a typical object must be roughly three
times greater than, or five times less than the minimum-energy value.

Observations of inverse-Compton emission in the X-ray have suggested
in a few well-studied cases that the field strengths are at or close
to equipartition with the radio-emitting electrons, both in the lobes
(e.g. Feigelson \etal\ 1995, Tsakiris \etal\ 1996) and hotspots
(3C\,405, Harris, Carilli \& Perley 1994; 3C\,295, Harris \etal\ 2000;
3C\,123, Hardcastle \etal\ in prep.) of radio galaxies. Field
strengths much lower than the equipartition value would give rise to
substantial inverse-Compton X-ray emission from the lobes in many
sources, but there are few detections to date. Having said this, there
are several sources (e.g.\ 3C\,120, Harris \etal\ 1999; Pictor A,
R\"oser \& Meisenheimer 1987) with X-ray hotspots which are too bright
to be consistent either with an X-ray synchrotron model (without
invoking a separate population of electrons) or with inverse-Compton
emission at equipartition, which may be evidence that field strengths
are far below equipartition in the hotspots of some sources. The X-ray
jet in the quasar PKS 0637-752 is a particularly dramatic case of
radio-related X-ray emission which cannot easily be explained using
the equipartition assumption (Chartas \etal\ 2000).

\subsection{Filling factors}

If the synchrotron-emitting plasma has a volume filling factor $\phi$
less than unity, then synchrotron volume emissivities are
underestimated, and the corresponding minimum pressures and energy
densities $u_{\rm TOT}$ in the plasma increase by a factor $\sim
\phi^{-4/7}$ (equation \ref{mine}). But the dynamically interesting
quantity is the energy density averaged over the volume of the lobe,
which is proportional to $\sim \phi^{3/7}$, so a filling factor $\phi
< 1$ cannot on its own account for the discrepancy between internal
and external pressures. If we assume (as is often implicitly done in
discussions of low filling factor) that a non-radiating `fluid' with
an energy density roughly equal to $u_{\rm TOT}$ fills the gaps
between emitting regions, then the total energy density increases as
$\sim \phi^{-4/7}$, and filling factors of $\phi \sim 0.02$ are
required to make the lobe pressures similar to the pressures in the
X-ray-emitting gas.

For a given synchrotron flux level, the observed X-ray flux from
inverse-Compton scattering of CMB photons is proportional to the mean
number density of electrons of appropriate energies in the lobes,
which is proportional to $\phi^{3/7}$, so lower filling factors would
mean lower inverse-Compton fluxes from the lobes. But if the
space-filling `fluid' is relativistic electrons, so that we have a
uniform electron population with strong field-strength variations, the
CMB inverse-Compton flux from the lobes should vary as $\phi^{-4/7}$,
since the number density of electrons outside the emitting regions is
roughly the same as that inside the emitting regions. On the other hand, the
flux from synchrotron-self-Compton emission (e.g.\ from hotspots) is
dependent on the number density of electrons in the {\it emitting}
regions, which is proportional to $\phi^{-4/7}$, so that we would in
general expect higher X-ray fluxes from synchroton-self-Compton
emission if the filling factor were low, although the actual flux is
strongly influenced by the geometry of the emitting region, which
affects the number density of photons available for scattering. If
the space-filling fluid is electrons, the geometry dependence of this
process is weaker.

For $\phi \sim 0.02$ throughout the source, we would thus expect X-ray
emission from inverse-Compton scattering of CMB photons in the lobes to be
roughly a factor of 5 less than, or a factor of 10 greater than the
equipartition predictions, while synchrotron-self-Compton emission
from hotspots should be higher than predicted by a factor of 10; although
there are few existing observations, the data (section \ref{icobs})
suggest that the filling factor is not this low. But because the
dependences on $\phi$ are so weak, we cannot rule out a
contribution from low filling factors to the pressure discrepancy.

\subsection{Non-radiating particles}
\label{nonr}

A large contribution to the energy density in the lobes may be made by
protons and other particles, such as low-energy electrons, which do
not emit synchrotron radiation in observable wavebands. Firstly, the
lobes may contain thermal protons. The lack of internal Faraday
depolarization of radio lobes places some limits on the internal
thermal particle content given simple models for the magnetic field
structure (e.g.\ Dreher \etal\ 1987), but large amounts of thermal
material can be hidden by field reversals (Laing 1984) so these limits
are not generally very useful. In FRI radio galaxies there is almost
certainly some contribution to the energy density from thermal protons
entrained by the trans-sonic jets, but this cannot account for the
whole of the pressure discrepancy even in those objects, because some
well-studied sources (e.g. B\"ohringer \etal\ 1993; Hardcastle,
Worrall \& Birkinshaw 1998) show X-ray deficits in the lobes which
would not be observed if a large amount of thermal gas at a
temperature comparable to that of the external medium was present in
them. (Cooler gas could be present but proportionally higher densities
would be needed to provide a useful contribution to the internal
pressure; it is hard to see where very hot thermal protons would come
from, as there is no obvious efficient heating process.) Deficits of
X-ray emission associated with the lobes are also reported in the FRII
Cygnus A (Carilli, Perley \& Harris 1994), which suggests that thermal
material cannot account for the pressure discrepancy in this source
either. In any case, the supersonic jets of FRIIs have less
opportunity to entrain thermal material, though they may pick up some
mass from stellar winds as they pass through the galaxy (Bowman, Leahy
\& Komissarov 1996).

Alternatively, FRII jets may be electron-proton from the start, with
relativistic protons providing the additional pressure required. From equation
\ref{mine}, the minimum pressure is proportional to $(1 +
\kappa)^{4/7}$, where $\kappa$ is the ratio between the numbers of
non-emitting and synchrotron-emitting particles and the two
populations of particles are assumed to have the same energy
distribution. Values of $\kappa \sim 60$ are thus required to increase
the pressure by a factor 10. However, equation \ref{mine} applies for
equal energy densities in particles and magnetic field. If (as the
measurements of inverse-Compton emission would indicate) the equality
is between magnetic field energy density and energy density in {\it
radiating} particles only, the energy density in relativistic protons
need only be $\sim 20$ times that in electrons or magnetic field to
bring the lobes back into pressure balance with the external
medium. As discussed by Leahy \& Gizani (1999), this still has a large
effect on the energy requirements for the source. Arguments for a
proton-dominated jet are presented by e.g.\ Celotti \& Fabian (1993).

Finally, it is possible to hide some energy density in electrons and
positrons with low energies (Lorentz factors $< 10$) which, for
magnetic field strengths around equipartition, do not radiate in
observable wavebands in the lobes. For our standard low-energy
electron energy index of 2, extending our assumed low-energy cutoff of
$\gamma_{\rm min} = 10$ down to $\gamma_{\rm min} = 1$ does not help a
great deal, increasing the equipartition energy density by only $\sim
10$ per cent; the effects would be greater (equation \ref{mine}) if
the energy index were steeper. But if there is a large
sub-relativistic population of particles which do not follow the
power-law distribution in energy, it is possible to make a substantial
difference to the energy density.  It has been argued that
$\gamma_{\rm min} \ga 100$ is required in the bases of AGN jets to
reproduce the observed levels of synchrotron self-Compton emission
(Ghisellini \etal\ 1992), though this argument is somewhat sensitive
to the details of energy transport close to the nucleus. If
$\gamma_{\rm min} > 100$ in the jets then we do not expect to see a
substantial population of $\gamma \ll 100$ electrons in the lobes (the
timescales for synchrotron/IC loss seem too long for these processes
to produce a significant low-$\gamma$ population of electrons). If
Wardle \etal\ (1998) are correct, however, low-energy ($\gamma \sim
1$) electrons are required to provide the Faraday conversion giving
rise to circular polarization in the radio jets of quasars, and so
there is some scope remaining for accounting for some of the missing
pressure in this way.

\section{Consequences for models}
\label{conseq}

FRII radio sources cannot be underpressured with respect to the
external medium for the whole of their length, as the minimum
pressures would suggest, or we would not observe lobes at all.  The
data force us, like Leahy \& Gizani (1999), to the conclusion that
there is some additional contribution to the internal pressure in at
least some, and maybe all FRIIs. The most likely candidates, from the
discussion above, are internal protons, magnetic field strengths a
factor of a few away from the minimum-energy values, or low filling
factors. All of these require coincidences to explain the similarity
of the magnetic field strengths derived from observations of
inverse-Compton emission to the minimum-energy values in a few
sources, but are otherwise consistent with observation.
As yet there are few measurements of inverse-Compton emission from the
lobes of powerful FRIIs due to the difficulty of detecting
low-surface-brightness extended X-ray features and distinguishing them
from the X-ray-emitting atmospheres; but see Tsakiris \etal\ (1996)
for observations of a few low-$z$ giant objects which suggest field
strengths close to, but slightly below, the equipartition value with
no proton contribution. If further observations confirm that field
strengths are normally close to the levels predicted by equipartition
arguments, then the internal-proton model seems the best contender,
since it can most easily accommodate such a result.

However, our observations provide no support for the common assumption
that radio sources are highly {\it overpressured} over their whole
length with respect to the external medium, as in model A of Scheuer
(1974). To produce highly overpressured sources, with supersonic
lateral expansion, we would need a still larger contribution from
protons, low filling factors or non-equipartition field strengths in
the lobes. The X-ray data do not rule out a model for radio-source
dynamics more similar to Scheuer's model C, in which the sources have
lobe pressures comparable to the external pressure in the X-ray
atmosphere, at least by the time they reach linear sizes of hundreds
of kpc. Indeed, from the point of view of the power that is required
to be transported by the jet, such a model is the most parsimonious we
can construct.

This has implications for the self-similar models described by Kaiser
\& Alexander (1997, hereafter KA), which require radio sources to be
described by model A. In these models (and indeed in any realistic
model of a radio source) the lobe pressure decreases as a function of
time or source length. So it is quite possible for a source to start
off highly overpressured (as it seems the small sources discussed in
section \ref{small} must be), and later to come into equilibrium with the
external medium. From equations 31 and 34 of KA, it can be seen that
for a source of constant jet power there is an approximately linear
decrease in lobe pressure with source length given some simple
assumptions about the external atmosphere\footnote{Williams (1991)
derives a stronger but qualitatively similar dependence on source
length on the assumption of a uniform external medium.}. As discussed
in section \ref{intro}, because the pressure in the external medium
decreases with distance from the cluster centre, which is in general
coincident with the central nucleus of the radio source, while the
pressure in the lobes is constant at a given moment along the source
length because of the high internal sound speed, a source can be
underpressured in its inner regions while being overpressured (and
continuing to expand transversely) further from the nucleus. The inner
parts of the lobes will be crushed by the external thermal medium on a
timescale given by the sound crossing time in the medium, which is
typically of the order $10^8$ years (comparable to the lifetime of the
radio source).  The result will therefore be a slow contraction of the
inner lobe [the `cocoon crushing' of Williams (1991)] eventually
removing the radio-emitting plasma altogether from the central regions
of the source. Although not fatal to the source, the contraction of
its inner regions will involve a departure from self-similarity.
\label{crush}

There may be some evidence for this process in the tapered and
sometimes absent inner lobes seen in some FRII sources (although this
may to some extent just be a result of spectral ageing), and the
compact appearance of high-redshift, luminous radio sources (Jenkins
\& McEllin 1977) may be a result of their rich cluster
environments. Hardcastle (1999) speculated that cocoon crushing might
even account for the appearance of wide-angle-tail radio sources in
clusters. Observations suggesting that axial ratio is independent of
length may be failing to take account of the variation of width along
the source, and studies of large samples of sources do show a weak
correlation of axial ratio with length (Black 1992),
subject to the same caveat about the effects of spectral ageing. So it
seems that both the X-ray and radio data suggest the possibility of a
breakdown of self-similar expansion, at least in older sources.

To make this argument quantitative, we have applied the KA model to
some of the sources in Tables \ref{ss} and \ref{limits}. Our X-ray
data provide the necessary information on the cluster density as a
function of radius; together with the jet power and the source lengths
and axial ratios (measured from radio maps) they allow us to calculate
the expected internal pressures in the lobes. If we use the widely
adopted values of jet power $Q_0$ calculated by Rawlings \& Saunders
(1991), which are based on minimum-energy assumptions, we compute
expected lobe pressures which are in good agreement with the minimum
pressures we have derived for the sources in all cases where the
approximations we use (chiefly those involved in mapping our
$\beta$-models onto KA's simplified density profiles) are
applicable. But the computed pressures, like the minimum pressures,
lie well below the external thermal pressure, whereas in the KA model
the internal pressure must always be above the external pressure. The
KA model thus {\it cannot} consistently describe these radio sources
if the jet powers are as low as those estimated by Rawlings \&
Saunders, unless the estimated cluster temperatures are much too high;
equivalently, we could say that the KA model together with minimum
energy assumptions `predicts' cluster temperatures of order 0.1 keV,
much lower than observed. In the KA model, the internal pressure
scales as $Q_0^{2/3}$ for a radio source of fixed length in a fixed
environment, so we need large increases in $Q_0$, by 2--3 orders of
magnitude, to produce sources which will be overpressured at the
cluster centre when the jet length is hundreds of kpc, as the KA model
requires for self-similarity. A smaller increase in $Q_0$ will produce
sources which at large jet lengths are susceptible to the
cocoon-crushing process described above, and so which deviate from the
self-similar model.

\section{Conclusions}

We have examined the {\it ROSAT} observations for all 3CRR FRIIs for
which pointed data exist, and estimated thermal pressures in their
X-ray emitting atmospheres. Although many of our estimates of thermal
pressure are limits, our data strongly suggest that many, and maybe
most, FRIIs with linear sizes $\ga 10$ kpc have lobe minimum pressures
which lie below the external thermal pressure.  FRIIs are thus
probably similar to the better-studied population of FRIs. Since it is
not physically possible for lobes to be so strongly underpressured,
the implication is that one or more of the standard minimum-energy
assumptions is wrong. The most obvious way of solving this problem is
to have a dominant contribution to the energy density in the lobes
from non-radiating particles such as protons, though we cannot rule
out other possibilities, such as a low volume filling factor for the
radio-emitting plasma. It is then a `coincidence' that a few studies
have found magnetic field energy densities in lobes and hotspots close
to the energy density in electrons. Further observations with {\it
XMM} and {\it Chandra}, which should allow the routine detection of
inverse-Compton emission from the lobes and hotspots of radio sources,
will allow these possibilities to be tested in more detail.

The transverse expansion of FRII sources will remain supersonic over
their expected lifetimes, as in model A of Scheuer (1974), and
therefore self-similar, as in the models of KA, only if the internal
pressures (and consequently the power supplied by the jet, $Q_0$) are
typically several orders of magnitude above the minimum values. In
more parsimonious models, with values of $Q_0$ only an order of
magnitude above the minimum-energy values quoted by Rawlings \&
Saunders (1991), the expansion of sources can initially be
self-similar but there will be departures from self-similarity for
large objects. Without an independent way of estimating the proton
content of lobes we have no way of knowing which of these situations
really obtains, but purely on energy budget grounds we feel that
modellers should be reluctant to rule out the low-$Q_0$ scenario. At
least for large sources, model C of Scheuer (1974) may be the right
one to use after all.

\section*{Acknowledgements}

We are grateful to Christian Kaiser for helpful comments on an early
draft of this paper, to Mark Birkinshaw for comments on a later
version, and to Guy Pooley and George Gilbert for providing electronic
radio maps of high-$z$ radio galaxies. We
thank an anonymous referee for comments which enabled us to improve
the paper.  This work was supported by PPARC grant GR/K98582, and made
use of NASA's Astrophysics Data System Abstract Service.

\bsp

\end{document}